# VoIP over Multiple IEEE 802.11 Wireless LANs

An Chan, *Graduate Student Member, IEEE,* Soung Chang Liew, *Senior Member, IEEE*

**Abstract**— IEEE 802.11 WLAN has high data rate (e.g., 11Mbps for 802.11b and 54Mbps for 802.11g) while voice stream of VoIP has low data rate requirement (e.g., 29.2Kbps). One may therefore expect WLAN to be able to support a large number of VoIP sessions (e.g., 200 and 900 sessions in 802.11b and 802.11g, respectively). Prior work by one of the authors, however, indicated that 802.11 is extremely inefficient for VoIP transport. Only 12 and 60 VoIP sessions can be supported in an 802.11b and an 802.11g WLAN, respectively. This paper shows that the bad news does not stop there. When there are multiple WLANs in the vicinity of each other – a common situation these days – the already-low VoIP capacity can be further eroded in a significant manner. For example, in a 5-by-5, 25-cell multi-WLAN network, the VoIP capacities for 802.11b and 802.11g are only 1.63 and 10.34 sessions per AP, respectively. This paper investigates several solutions to improve the VoIP capacity. Based on a conflict graph model, we propose a clique-analytical call-admission scheme, which increases the VoIP capacity by 52% from 1.63 to 2.48 sessions per AP in 802.11b. For 11g, the call-admission scheme can also increase the capacity by 37%, from 10.34 to 14.14 sessions per AP. If all the three orthogonal frequency channels available in 11b and 11g are used to reduce interferences among adjacent WLANs, clique-analytical call admission can boost the capacity to 7.39 VoIP sessions per AP in 11b and to 44.91 sessions per AP in 11g. Last but not least, this paper expounds for the first time the use of coarse-grained time-division multiple access (CoTDMA) in conjunction with the basic 802.11 CSMA to eliminate the performance-degrading exposed-node and hidden-node problems in 802.11. A 2-layer coloring problem (which is distinct from the classical graph coloring problem) is formulated to assign coarse time slots and frequency channels to VoIP sessions, taking into account the intricacies of the carrier-sensing operation of 802.11. We find that CoTDMA can further increase the VoIP capacity in the multi-WLAN scenario by an additional 35% to 10 and 58 sessions per AP in 802.11b and 802.11g, respectively.

—————————— ◆ ——————————

## 1 INTRODUCTION

VOICE-OVER-IP (VoIP) is one of the fastest growing applications for the Internet today. At the same time, driven by huge demands for portable access, the market for wireless Local Area Network (WLAN) based on the IEEE 802.11 standard is taking off quickly. Many cities are planning on city-wide deployment of WLAN. An important application over these networks will be VoIP over WLAN.

A hurdle, however, is the low number of voice conversations that can be supported. As shown in previous investigations [1, 2] by one of the authors, although in theory many voice sessions can be supported in an 802.11b WLAN based on simplistic raw-bandwidth calculation, in reality only less than 12 can be accommodated. There has been much subsequent work on how to improve the quality-of-service (QoS) of VoIP over WLAN. Part of the IEEE 802.11e standard [3], for example, is to address that.

Most of the prior investigative efforts [1, 2, 4-6], have been focused on the single isolated WLAN scenario. In practice, with the proliferation of WLAN these days, it is common to find numerous WLANs within a small geographical area – one only needs to do a cursory scan with a WLAN-equipped personal computer to see the considerable number of WLANs within a building. Although recently there has been much attention paid to multihop wireless mesh networks and VoIP over such networks [7-9], infrastructure WLAN is still the most widely deployed architecture nowadays. This paper is a first attempt to examine the VoIP capacity in the "multi-cell" environment in which many WLANs are deployed in the same geographical area.

We find that the VoIP capacity is further eroded in the multi-cell scenario, and substantially so. For example, our NS2 [10] simulations show that the capacity of a 5-by-5, 25-cell WLAN is only 1.63 VoIP sessions per access point (AP) in 802.11b, and 10.34 sessions per AP in 802.11g. This dismal performance has important implications that deserve further attention in view of the accelerating productization of VoIP-over-WLAN technologies.

Besides pointing out the alarmingly low efficiency of VoIP over multi-cell WLAN, and identifying the mutual interferences of the CSMA operation of adjacent cells as the major culprit for the dismal performance, this paper is also a first foray into finding solutions to alleviate the problem. Based on a conflict-graph model, we set up a framework for call admission control so as to better man-

- A. Chan graduated from the Department of Information Engineering, The Chinese University of Hong Kong. He is now studying Ph.D program in the Department of Computer Scinece in the University of California, Davis, CA 95616. E-mail: achan5@ ie.cuhk.edu.hk.
- S.C. Liew is with the Department of Information Engineering, The Chinese Univieristy of Hong Kong, Hong Kong. E-mail: soung@ie.cuhk.edu.hk.

age the mutual interferences. The simulation results show that a clique-analytical call-admission scheme can increase the VoIP capacity to 2.48 VoIP sessions per AP in 802.11b (i.e., 52.1% improvement) and to 14.14 sessions per AP in 802.11g (i.e., 36.75% improvement) for the 5-by-5, 25-cell WLAN. If all the three orthogonal frequency channels in 802.11b/g are used, the clique-analytical call admission control can boost the per-AP capacity to 7.39 VoIP sessions in 802.11b and to 44.91 sessions in 802.11g.

Another major contribution of this paper is the proposal of a coarse-grained time-division multiple-access (CoTDMA) approach to alleviate the multi-cell mutual interferences. In CoTDMA, the time dimension is divided into multiple coarse time slots. Multiple VoIP sessions are then assigned to each time slot, and they make use of the basic 802.11 CSMA protocol to coordinate channel access within the time slot. Coarse-grained time slots could be implemented using the sleep mode of 802.11, originally intended for power conservation purposes. The basic idea of CoTDMA is that VoIP sessions of adjacent WLANs that interfere with each other should be assigned different time slots, so that VoIP sessions of *different* cells do not need to use CSMA to coordinate transmissions among them; essentially CSMA needs to be effective only among the sessions of the *same* cell.

As will be shown in this paper, the theoretical call-admission control framework of CoTDMA corresponds to a new class of graph-coloring problem that is distinct from that of the classical graph-coloring problem. With only three coarse-grained time slots, VoIP capacity per AP can be boosted to 10 and 58 sessions in 802.11b and 802.11g, respectively. This is another 35.3% improvement over the three-frequency-channel clique-analytical call admission control strategy.

The remainder of the paper is organized as follows. Section 2 explains how the 802.11 CSMA protocol would affect the VoIP capacity. In particular, we show that the VoIP capacity can be eroded further in a significant way in the multi-cell setting. Section 3 presents a call admission strategy based on clique analysis of a graph-theoretic formulation to confine inter-cell interferences. Section 4 considers using a time-dimension approach in conjunction with the basic 802.11 CSMA to further improve the VoIP capacity. Section 5 concludes the paper.

## 2 LOW VOIP CAPACITY OVER MULTIPLE WLANS

### 2.1 802.11 Protocols and VoIP

VoIP packets are streams of packets containing encoded voice signals. There are different codecs for encoding voice signals. Take GSM 6.10 as an example. The voice payload is 33-byte and 50 packets are generated in each second. After adding the 40-byte IP/UDP/RTP header, the minimum channel capacity to support a GSM 6.10 voice stream in one direction (either uplink or downlink) is 29.2Kbps.

An 802.11b WLAN in theory can support nearly 200 VoIP sessions (divide 11Mbps by two times of 29.2Kbps); and for 802.11g, more than 900 sessions (divide 54 Mbps by two times of 29.2Kbps). However, prior investigation has shown that the actual VoIP capacity is severely limited due to various inherent header and protocol overheads. With the GSM 6.10 codec, for example, only 12 VoIP sessions can be supported in an 802.11b WLAN; and 60 sessions in 802.11g [1, 2].

Prior work has primarily focused on VoIP over an isolated WLAN. When several WLANs are in the proximity of each other, their VoIP sessions may compete with each other for airtime usage if these WLANs use the same frequency channels. In 802.11b/g, for example, there are only three orthogonal channels, but it is not uncommon to see more than three overlapping WLANs in a building these days. We show in this paper that the already-low efficiency of VoIP over a single WLAN will be further eroded in such a situation, so much so that only an average of less than two VoIP sessions per AP can be supported in a 5-by-5, 25-cell, 802.11b WLANs; and around 10 sessions in 802.11g.

### 2.2 Low VoIP Capacity over Multiple WLANs

Let us assume the allowance of packet-loss rate of 1% to 3%. Then, the minimum channel capacity requirement is 28.32Kbps for GSM 6.10 codec. If both the uplink and downlink of a VoIP session can have throughputs exceeding this benchmark, we say that the VoIP session can be supported in the WLAN.

To evaluate the VoIP capacity over multiple WLANs, we model a "WLAN cell" with a regular hexagonal area of 250m on each side. An AP is placed at the center of the cell. Any wireless client station inside the cell will be associated with the AP. The longest link distance ($d_{max}$) is therefore 250m, which is the data transmission range (*TXRange*) for 802.11b assumed in NS2. The transmission ranges of APs in WLANs partially overlaps. This is also the case in practice. The circles in Fig. 1 represent the coverage of certain WLANs (transmission ranges of APs). In this paper, instead of using circles to model WLAN cell, we use hexagons for clearer illustration. The results we derived from this hexagonal-area model can be directly applied to the circular-area model. By placing the cells side by side, we form a *D*-by-*D* multi-cell topology, where *D* is the number of cells on each side. Fig. 1 shows a 3-by-3 multi-cell topology.

We consider the use of the basic mode [11] of 802.11 in this paper because the short VoIP payload does not warrant the use of RTS/CTS. We assume the carrier sensing range (*CSRange*) of all wireless stations is 550m, the default in NS2.

A point to note is that in real equipment, the operating parameters are based on "power thresholds" rather than "distances". For example, for carrier sensing and for detection, it is the power received that matters rather than the distance, and there is a one-to-one correspondence between distance and power only if the wireless propagation medium is spatially homogeneous. The implementations of the schemes (based on a graph-theoretic formulation) expounded in this paper are compatible with the power-threshold interpretation and that the assumption of a homogeneous medium is not necessary. Nevertheless, for convenience we will continue to use the distances (e.g. *TXRange* and *CSRange*) to describe the system operation. Thus, "distance" is to be interpreted in a virtual sense, explained as follows. Two wireless stations $i$ and $j$ are said to be separated by a virtual distance $d_{i,j}$ if the power they receive from each other is $P_{i,j} = P_{j,i} = k/d_{i,j}^\alpha$, where $k$ is a constant, $\alpha$ is a "reference" (not the actual) path-loss exponent, assuming all stations use the same transmission power. Applying common $k$ and reference to the whole network, we can then derive the virtual distance $d_{i,j}$ from the measured power transferred, $P_{i,j}$. Given a power threshold requirement, there is then a corresponding virtual-distance requirement. When we say the *CSRange* is set to $d_{CS}$, we mean the carrier-sensing threshold power is set to $P_{CS} = k/d_{CS}^\alpha$ where $\alpha$ is the nominal constant adopted (e.g., $\alpha = 4$). In the subsequent discussions, we assume there is an underlying scheme to find out $P_{i,j}$, hence, $d_{i,j}$, so that we can assign system resources (e.g., AP association and time-slot assignment) according to $d_{i,j}$. To limit our scope, however, we will not discuss the $d_{i,j}$ discovery algorithms here. The reader is referred to [12] for possible schemes. In short, the logical correctness of the implementation of our schemes in this paper does not depend on spatial homogeneity. The performance, however, does depend on the $\alpha$ being assumed.

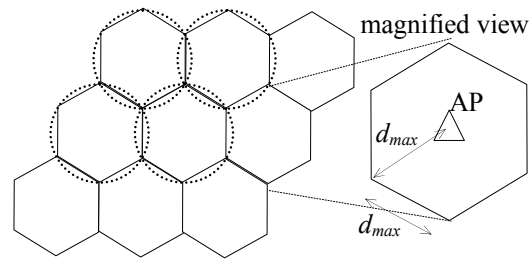

Fig. 1. A 3-by-3 multi-cell topology.

We ran simulation experiments on the *D*-by-*D* multi-cell topology for $D = 1, 2, 3, 4$, and 5. In each run, wireless stations (VoIP sessions) are added one by one randomly assuming uniform distribution. When a particular cell has 12 stations in the 802.11b case (60 stations in the 802.11g case), no more stations will be added to that cell to ensure the capacity of the single isolated WLAN case is not exceeded. With each additional VoIP session, NS2 simulation is run and the throughput of each link recorded. When the next newly added VoIP session causes violation of the packet-loss rate requirement (1% to 3%) by at least one of the sessions, we say that the capacity limit has been exceeded. This corresponds to a simplistic call admission scheme in which upon unacceptable performance caused by the newly added session, the newly added session will be dropped, and no more future sessions will be accepted. We will later consider a cleverer call admission scheme based on a clique analysis of a conflict graph so that we can "predict" the performance before deciding whether to admit a call to get rid of the disruptiveness of adding a session only to drop it later. Table 1 summarizes our simulation results.

In Table 1, $C_{D \times D}$ is the total number of VoIP sessions that can be supported in a *D*-by-*D* multi-cell topology. Obviously, as *D* increases, more VoIP sessions can be supported. We further calculate $C_{AP\_D}$, the per-AP capacity in a *D*-by-*D* multi-cell WLAN, defined as follows:
$$C_{AP\_D} = C_{D \times D} / D^2 \quad (1)$$

We plot $C_{AP\_D}$ against number of cells, $D^2$, in Fig. 2. We find that as the number of cells increases, per-AP capacity decreases. When the number of cells is 25, i.e. $D = 5$, only 1.63 VoIP sessions can be supported by each AP in 802.11b. Compared with the single-cell scenario, where each AP can support 12 VoIP sessions in 802.11b, this is a rather large penalty! Similar capacity penalty is also observed in 802.11g, where only 10.34 sessions per AP can be supported when $D = 5$, as opposed to 55 in the single-isolated WLAN case when $D = 1$.[1]

We also note in passing that for a network larger than the 5-by-5 network, most cells will be surrounded by six adjacent cells and there will be proportionately fewer cells at the boundary, where they enjoy less interference

TABLE 1

VOIP CAPACITY OVER *D*-BY-*D* MULTI-CELL TOPOLOGIES

| D | 1 | 2 | 3 | 4 | 5 |
|---|---|---|---|---|---|
| Avg. $C_{DxD}$ in 11b | 12.0 | 12.3 | 20.0 | 30.0 | 40.8 |
| Avg. $C_{DxD}$ in 11g | 55.0 | 79.6 | 131.8 | 207.6 | 258.4 |

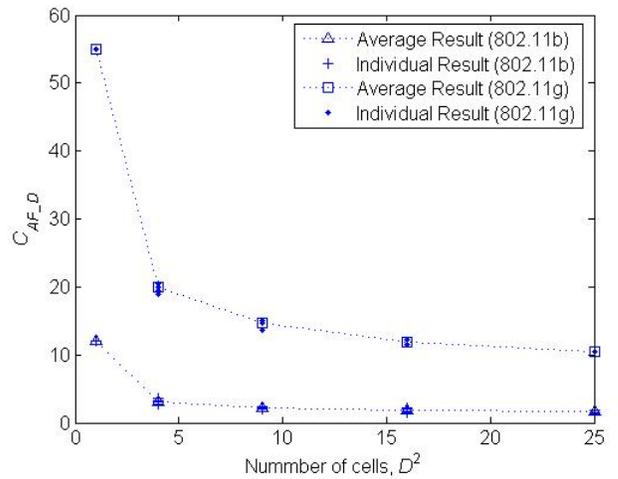

Fig. 2. Per-AP capacity of *D*-by-*D* multi-cell WLAN.

---

[1] The VoIP capacity of a single-cell 802.11g WLAN measured in simulation is 55, not 60 as calculated in [1, 2]. This is due to the smaller minimum contention window size in 802.11g. But for consistency of comparison, we still take "60 VoIP sessions" as the theoretical upper bound for VoIP capacity over single isolated 802.11g WLAN in our following discussions

from other cells. One can therefore expect the VoIP capacity per AP to drop even further when the dimensions are larger than 5-by-5. Indeed, in a 10-by-10, 100-cell, 802.11b WLAN, our simulation shows that the per-AP capacity is only 1.22 sessions.

## 2.3 Applying Frequency-Channel Assignment

To reduce the mutual interference of neighboring cells, a quick solution is to assign different frequency channels to different cells, as in cellular telephone networks. If there are enough frequency channels, the neighbor cells that could interfere with each other through packet collisions and carrier sensing could be assigned different frequency channels. This boils down to the same situation as in the single-cell case, so that the per-AP VoIP capacity in the multi-cell case is the same as that in the single-cell case.

IEEE 802.11b/g has only three orthogonal frequency channels, and this is not sufficient to completely isolate co-channel interference between cells. Fig. 3 shows that if we have only three frequency channels, then the nearest distance between two cells using the same channel (e.g. the two un-shaded hexagons in Fig. 3) is the same as the maximum link length within a cell, $d_{max}$. Since these two cells may interfere with each other, the carrier-sensing range (*CSRange*) should be larger than $3d_{max}$ to avoid hidden-node [13] collisions between the two cells. To see this, consider link (AP1, STA1) and link (AP2, STA2) of the two cells in Fig. 3, where the distance between STA1 and STA2 is $d_{max}$. The transmission of STA1 will interfere with the reception of STA2, and vice versa, so the two APs should be able to carrier sense each other for not starting transmission on their own links simultaneously. Hence, *CSRange* of $3d_{max}$ is needed to avoid hidden-node collisions between these two links. Since a cell must now share airtime with other cells, the VoIP capacity per AP cannot be the same as that in the single-cell case.

802.11a, on the other hand, has twelve orthogonal channels. Fig. 4 shows that a 7-channel assignment is sufficient for complete isolation of co-channel interference. The nearest distance between two cells using the same channel is $2.65d_{max}$ which is larger than the minimum *CSRange*, $2d_{max}$, used to prevent collisions *within a cell*. Thus, with 7-channel assignment, co-channel interference can be completely isolated. However, if we simply use 7 overlay networks in each cell (put 7 APs inside each cell and operate in different channels), the number of VoIP sessions that can be supported in each cell can be increased by 7 times of that in the single-channel multi-cell topology. Therefore, we find that the channel assignment in Fig. 4 actually may not improve the VoIP capacity on a per-frequency channel basis, although on a per-AP basis, it does. Furthermore, IEEE 802.11a operates in regulated frequency band (license is needed), so it is not commonly deployed.

In short, the case of 802.11b/g in which there are not sufficient frequency channels for complete elimination of co-channel interference among adjacent cells will be of much practical interest. Section 3 considers how to perform call admission within each frequency plane in which cells assigned with the same frequency channel may mutually interfere with each other.

## 3 CLIQUE ANALYSIS AND CALL ADMISSION

To understand the cause for the heavy performance penalty in the multi-cell scenario, we consider here a clique analysis based on a graph model that captures the conflict and interference among the nodes. The clique analysis also suggests a call admission methodology. With this call admission scheme, the VoIP capacity can be increased to 2.48 sessions per AP from 1.63 in the case of 5-by-5, 25-cell 802.11b WLAN. This constitutes a 52.1% improvement. In the 5-by-5, 25-cell 802.11g WLAN, this call admission scheme increases the per-AP capacity by 36.75% from 10.34 to 14.14 sessions.

### 3.1 Conflict Graphs and Cliques

In our conflict graph model, vertices represent VoIP sessions (wireless links). An edge between two vertices means that the two VoIP sessions compete for the airtime. In other words, they cannot transmit at the same time. There are two reasons why they cannot transmit at the same time. (i) First, nodes of the two sessions that are within the *CSRange* of each other will be prevented by the 802.11 protocol from transmitting together. (ii) Even if the two sessions are not within each other's *CSRange*, there may be mutual interference between them so that either one or both of their transmissions will fail if they transmit together: this is due to the well-known hidden-node problem [13]. In either case (i) or (ii), an edge is drawn between the two vertices of the VoIP sessions to mean that their transmissions cannot use the same airtime. A clique is a subset of vertices in which there is an edge between any of two vertices [14]. The vertices in a clique compete for common airtime. That is, the sum of the fractions of airtimes used by the vertices should not exceed one.

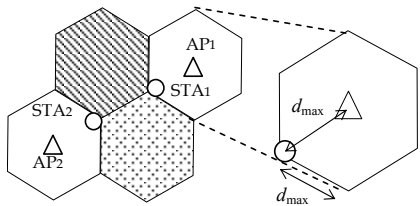

Fig. 3. 3-channel assignment in multi-cell WLAN.

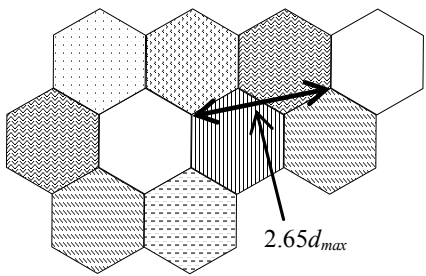

Fig. 4. 7-channel assignment in multi-cell WLAN.

In the single-cell scenario, all client stations are within the same cell and associated with the same AP. So, edges should be drawn among all vertices of the same cell. In an 802.11b single-cell WLAN, the maximum clique size is 12, because 12 VoIP sessions will fill up all airtimes [1, 2]. In an 802.11g single-cell WLAN, the maximum clique size is 60.

In a multi-cell WLAN, instead of one clique, multiple cliques can be formed. To see this, we need to inspect cases (i) and (ii) mentioned above carefully. For case (i), *CSRange* is usually a fixed value (assuming no power control). By default, it is 550m in NS2. For case (ii), we consider the Interference Range (IR) defined as follows:

$$IR_k = (1+\Delta)d_k \quad (2)$$

where $IR_k$ is the *Interference Range* of a node, $k$, (it can be a client station or an AP), $d_k$ is the length of the link associated with the node $k$; and $\Delta$ is a distance margin for interference-free reception with typical value of 0.78 [10, 13]. Within a radius of $IR_k$, any other transmission will interfere with the node $k$'s reception of the signal. The maximum link length within a cell is $d_{max} = 250m$ (shown in Fig. 1). The corresponding maximum IR is therefore $IR_{max} = 1.78 \times 250m = 445m$.

In Fig. 5, the distance between any two points of two different shaded areas is at least 866m, which is larger than both *CSRange* and $IR_{max}$. Therefore, there is no edge between vertices of different shaded cells, and the vertices of different shaded cells belong to different cliques.

From the discussion above, we know that there is a maximum clique size (e.g., 12 and 60 for 802.11b and 802.11g single-cell WLANs respectively) which cannot be exceeded if satisfactory performance of VoIP sessions is to be attained. To increase the VoIP capacity over multiple WLANs, we need to pack the VoIP sessions (vertices) efficiently. We discuss this call admission control in the next subsection.

### 3.2 Clique-Analytical Call Admission Control

The ability to predict whether a new VoIP session can be admitted without causing performance problem is important in call admission. Previous work [6, 15] showed that an additional VoIP session to a single-cell WLAN which already reaches the maximum capacity can degrade the performance of all other existing VoIP sessions. This is also the case in multi-cell WLANs. In the 802.11b 2-by-2 network (in which the maximum capacity is around 12.3 as shown in Table 1), for example, if there are 14 VoIP sessions, three of them cannot fulfill the loss-rate requirement. That means only 11 sessions can be supported with acceptable performance, as opposed to the 12 sessions that can be supported when there are exactly 12 sessions in WLAN.

We consider a call admission control mechanism based on the clique size of the conflict graph.

Let $E_{v_i}$ be the set of vertices with which vertex $v_i$ has an edge. Let $K_{v_i}$ be the set of all cliques $C_{v_i,x}$ to which $v_i$ belongs, where $x = 1,2,..., |K_{v_i}|$ is the index of the cliques, and $|K_{v_i}|$ is the total number of cliques in $K_{v_i}$. Any pair of cliques in $K_{v_i}$ must satisfy (3) below so that all cliques in $K_{v_i}$ are "maximal" and not contained in another clique [16]. As defined in (4), $m_{v_i}$ is the size of the largest clique in $K_{v_i}$.

$$C_{v_i,x} \not\subset C_{v_i,y} \quad , \quad x \neq y, \ 1 \leq x,y \leq |K_{v_i}| \quad (3)$$

$$m_{v_i} = \max_x |C_{v_i,x}| \quad (4)$$

Fig. 6 gives an example of a conflict graph, where vertex $v_1$ has the following parameters.

$E_{v_1} = \{v_2, v_3, v_4, v_5\}$

$K_{v_1} = \{v_1, v_2, v_3, v_5\}, \{v_1, v_3, v_4\}, \ |K_{v_1}| = 2$

i.e., $C_{v_1,1} = \{v_1, v_2, v_3, v_5\}, C_{v_1,2} = \{v_1, v_3, v_4\}$

$m_{v_1} = 4$

The pseudo code of the admission control algorithm is given in Algorithm I. There are three procedures in the algorithm: Procedures A, B and C. When there is a new call request (i.e., a new vertex $v_i$ (VoIP session) wants to be added), Procedure A is first executed, wherein a copy of the state $(K_{v_j}, m_{v_j}) \ \forall \ v_j$ is first saved in case the admission of $v_i$ fails and we need to revert to the original state later. After that, Procedure B is executed. Procedure B updates the state $(K_{v_j}, m_{v_j})$ assuming the addition of $v_i$. To satisfy (3), a function call, $NO\_REDUNDANCY(K_{v_j})$

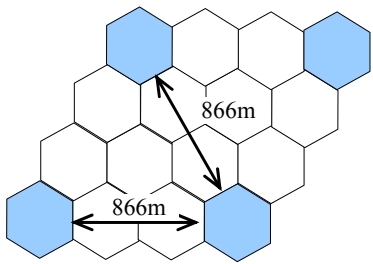

Fig. 5. . Multiple cliques may be formed in a 4-by-4 multi-cell WLAN

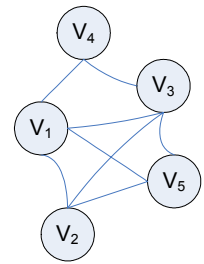

Fig. 6. An example of a conflict graph.

```
Algorithm I

Procedure A
Keep a copy of the state (K_{v_j}, m_{v_j}) for all existing v_j;
Perform Procedure B;

Procedure B
for each v_j ∈ E_{v_i} {
    for each C_{v_j,x} ∈ K_{v_j} {
        if C_{v_j,x} ⊂ E_{v_i}
            then add v_i to C_{v_j,x};
        else
            add {C_{v_j,x} ∩ E_{v_i}, v_i} to K_{v_j};
    }
    K_{v_j} = NO_REDUNDANCY(K_{v_j});
    update m_{v_j};

    if m_{v_j} > C_max,
        then reject v_i;
        revert the state using the copy stored in Procedure A;
        break out of Procedure B, and the algorithm is
        terminated;
} // All v_j ∈ E_{v_i} have been looked at if the algorithm comes to this point
Perform Procedure C;

Procedure C
v_i is admitted;
K_{v_i} = ∅;
for each v_j ∈ E_{v_i} {
    if v_i ∈ C_{v_j,x}
        then C_{v_i,x} = C_{v_j,x},
             K_{v_i} = K_{v_i} ∪ C_{v_i,x};
}
K_{v_i} = NO_REDUNDANCY(K_{v_i});
compute m_{v_i};
```

```
Algorithm II

NO_REDUNDANCY(K_{v_k}) {
    for each pair C_{v_k,x} ∈ K_{v_k}, C_{v_k,y} ∈ K_{v_k}, x ≠ y {
        if C_{v_k,x} ⊂ C_{v_k,y}
            then delete C_{v_k,x};
    }
    return (K_{v_k});
}
```

is made. Algorithm II gives the pseudo code of $NO\_REDUNDANCY(K_{v_k})$. During the updating, Procedure B continually checks to see if a pre-determined maximum clique size $C_{max}$ is exceeded so as not to violate the loss-rate requirement. If so, the algorithm is terminated and $v_i$ is rejected; in which case the state saved in Procedure A is restored.

If Procedure B successfully runs to the end without the $C_{max}$ being exceeded, Procedure C is executed. Procedure C admits the new vertex $v_i$ and calculates $(K_{v_i}, m_{v_i})$.

We have performed an experiment in MATLAB to measure the execution time of the call admission control algorithm. The experiment assumes the 5-by-5 multi-cell topology setting in Section 2.2. For simplicity, we first consider the single-frequency channel case in which all cells are assigned the same frequency. The next subsection will deal with the multi-frequency case.

We applied the algorithm to identify which VoIP sessions out of a total of 300 (12 x 25, and 802.11b is assumed here) randomly placed (with uniform distribution) potential sessions could be admitted in a 5-by-5 multi-cell WLAN with the $C_{max}$ clique-size constraint. Unlike the simplistic call admission scheme earlier, here when a session is rejected, the call admission scheme continues to consider a next session if the 300 sessions have not been exhausted. We ran several sets of experiments for different random node distributions in a 5-by-5 WLAN. We used an ordinary personal computer with 3.2GHz CPU and 1G RAM to perform the experiment. The results for $C_{max}$= 8 and 12 are shown in Table 2. The total runtime is the total time needed for the algorithm to go through all the 300 VoIP sessions. The average runtime is the time needed to admit or reject a call (Total runtime / 300). We see that although the general clique problem is NP-complete, the execution time of our algorithm on the conflict graph that models 802.11 networks is acceptable.

Based on the call admission results in MATLAB, we then used NS2 to verify whether the admitted calls meet the maximum 3% packet-loss rate requirement in the simulation. From Table 2, we find that when $C_{max}$ is 8, the average number of VoIP sessions admitted by our admission control algorithm is 62.0. This is the average number of five runs of the MATLAB experiments. We incorporated the corresponding sets of admitted VoIP sessions in five different runs of NS2 simulation. In each run, all the admitted VoIP sessions can meet our packet-loss rate requirement.

However, if instead of setting $C_{max}$ to 8, we set it to 12, then nearly one third of VoIP sessions cannot meet the loss requirement. It is interesting that for the large-scale multi-cell case, the maximum clique size that should be imposed is 8 rather than 12 (recall that 12 is the maximum clique size in the single-cell topology if 802.11b is assumed) if loss-rate requirement is to be satisfied. This is perhaps due to the interaction and "coupling" among different cliques caused by the 802.11 MAC protocol. In other words, 802.11 MAC may not achieve perfect scheduling in which the airtime usage within each clique is 100% tightly packed. This motivates us to explore the use of time-slot scheduling over the basic CSMA 802.11 MAC in Section 4 for performance improvement.

TABLE 2
EXPERIMENT RESULTS OF APPLYING THE ADMISSION CONTROL ALGORITHM ON 5-BY-5 MULTI-CELL WLAN

| $C_{max}$ | Total Runtime (s) | Average No. of Sessions Admitted | Average Runtime (s) |
|---|---|---|---|
| 8 | 48.6 | 62.0 | 0.160 |
| 12 | 112.1 | 70.4 | 0.374 |

With $C_{max}$ = 8, the 5-by-5 multi-cell WLAN can support 2.48 sessions (62.0/25) per AP, yielding a 52.1% improvement over the simplistic call admission scheme in Section 2. The clique-analytical call admission control works similarly well in 802.11g WLANs. For the 5-by-5 multi-cell 802.11g WLAN, with $C_{max}$ = 44, the per-AP capacity can be increased from 10.34 to 14.14 VoIP sessions, yielding a 36.75% improvement.

### 3.3 Clique-Analytical Admission Control in Three-Frequency-Channel WLANs

In this subsection, we explore the impact of multiple frequency channels on VoIP capacity over multiple WLANs. The availability of multiple frequency channels allows us to separate the cells using the same frequency by a longer distance. Farther separation of cells that use the same frequency leads to less conflict among transmissions of different VoIP sessions (although not eliminating conflicts entirely). Consequently, fewer edges are formed in the corresponding conflict graph.

Consider the multi-cell topology in Fig. 7, where we apply the three frequency-channel assignment (as in Fig. 3). In Fig. 7, the shaded cells use the same frequency channel, while the un-shaded cells use the other two frequency channels. Although the size of the topology in Fig. 7 is comparable to the 5-by-5, 25-cell WLAN we described in previous sections, the three frequency channels help to reduce conflicts and increase the number of VoIP sessions that can be supported by each AP. NS2 simulations show that applying the clique-analytical admission control to the three frequency-channel layout can boost the per-AP capacity to 7.39 VoIP sessions in 802.11b; and 44.91 VoIP sessions in 802.11g. These are respectively 2.98 and 3.17 times of the per-AP capacity in 802.11b and 802.11g 5-by-5, 25-cell WLAN where single-frequency channel is used.

In the next section, we explore time division on 802.11 MAC which can eliminate all HN and alleviate the exposed-node (EN) problem. The VoIP capacity over multiple WLAN can be further improved.

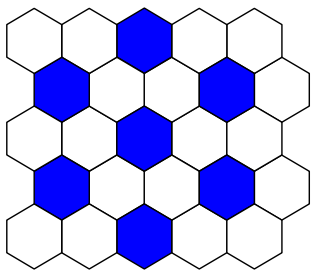

Fig. 7. 3-frequency-channel assignment is applied on multiple WLANs.

## 4 TIME-DIVISION CSMA MAC

So far we have learned that there is a significant capacity penalty when we move from the single-cell WLAN scenario to the multi-cell WLAN scenario. We have also learned that the extremely low VoIP capacity over multiple WLANs is due to the mutual interference from neighboring cells, and such mutual interference cannot be completely isolated even with careful frequency channel assignment. Although IEEE 802.11e has been standardized recently to support QoS in WLAN [3], it only focuses on the single-cell situation.

In this section, we explore adding the time-division approach to the basic 802.11 CSMA protocol. We show that the integrated time-division-CSMA approach can potentially increase the VoIP capacity over multiple WLANs quite significantly. Selected previous work has considered Time Division Multiple Access (TDMA) MAC. However, their focus is on the single WLAN case [17-20]. In addition, CSMA is proposed to be replaced entirely by TDMA [21-23] – i.e., the motivation was not to explore a solution workable within the context of the widely-deployed 802.11 technology.

For VoIP applications within 802.11, each transmission consists of a very small packet (relative to the raw data rate and the various packet headers), and "fine" TDMA requires tight synchronization among the stations, which can in turn cause throughput degradation. In this section, our primary focus is on the principle of "coarse" time division, in which a relatively large time slot is allocated to a group of stations. The stations of the same time slot then contend for channel access using the original 802.11 CSMA scheme. We (i) lay out and investigate a graph-theoretic problem formulation that captures the principle of integrating coarse time division with CSMA in Subsection 4.1; and (ii) provide a feasibility investigation of the approach within the context of 802.11 in Subsection 4.2.

### 4.1 Coarse-graind Time-Division Muliple Access

In Section 2.3, we have discussed 3-frequency-channel assignment in 802.11b/g WLAN (see Fig. 3) and argued that frequency-channel assignment alone cannot completely isolate co-channel interference. Section 3.3 applies the clique-analytical call admission control to boost per-AP VoIP capacity in 3-frequency-channel WLAN; however, the co-channel interference from different cells is still not isolated entirely. The goal of Coarse-grained Time-Division Multiple Access (CoTDMA) is to remove such co-channel interference. In particular, we impose a restriction such that two stations in different cells that interfere with each other would be allocated different time slots or different frequency channels.

*4.1.1 Basic Ideas of CoTDMA*

We first explain the concept using the *simplified* scenario depicted in Fig. 8. In Fig. 8, in addition to the 3-frequency-channel assignment, we divide each cell into six sectors and assign a distinct timeslot to each sector. The shaded cells use the same frequency channel, and the stations within each sector use the same time slot; the number in each sector indexes the time slot assigned to that particular sector. The frequency and time-slot assignments in Fig. 8 are such that different sectors do not interfere with each other because they are either active in different time slots, in cells of different frequencies, or sufficiently far apart from each other. This allows us to shorten the *CSRange* to $d_{max}$, the distance from the AP located at the center of each cell to the farthest corner in the cell. The nearest station with the same frequency and time-slot assignment in neighboring cells is at least $2d_{max}$ away, which is larger than both the *CSRange* and *IR* defined in (2). Therefore, the co-channel interference from neighboring cells is completely isolated. Note that within each sector (time slot), there may still be multiple stations, and the original 802.11 CSMA scheme is used to coordinate transmission among these stations. In the best-case scenario, the *VoIP capacity per AP* in multiple WLANs can be the same as that in the single isolated WLAN case. To see this, consider 802.11b: if each sector has two stations, then we can have a total of 12 VoIP sessions per AP.

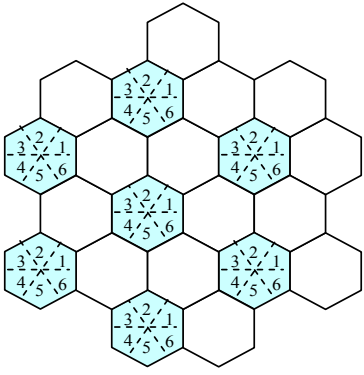

Fig. 8. 6-timeslot assignment in addition to 3-frequency-channel assignment in multiple WLANs.

Although the sectorized cells in Fig. 8 illustrate the concept of CoTDMA, it has two implementation difficulties: 1) time-slot assignment requires the knowledge of the locations of the individual stations; 2) if the stations are not evenly distributed across the sectors, then it will not be as effective as the best-case scenario mentioned above. In the following, we present a graph model of CoTDMA to solve these problems. The graph model presented also gives a framework for performance analysis of CoTDMA.

For convenience, we will continue to describe the system operation in terms of distances such as *CSRange* and *IR*. As stated in Section 2.2, implementation based strictly on the "geometric" distance interpretation is unnecessary once we move on to the graph-theoretic formulation here. A mapping of distances to power thresholds is all that is needed.

*Definition 1*: In CoTDMA, $m$ frequency channels and $n$ time slots are assigned to the VoIP sessions. In each cell, at most $k$ VoIP sessions are active in each time slot, where $k = \lfloor C_{AP\_1}/n \rfloor$, and $C_{AP\_1}$ is the average per-AP capacity in a single isolated WLAN.

### 4.1.2 Conflict Graph Modeling of CoTDMA

According to Definition 1, the parameter $m$ is the number of orthogonal frequency channels available. For example, $m = 3$ in 802.11b and 802.11g. If $n = 6$, then $k = 2$ in 802.11b, and $k = 10$ in 802.11g, since the respective $C_{AP\_1}$ are 12 and 60.

We will look at the system performance as a function of $n$ shortly. First, we formulate the corresponding graph-theoretic coloring problem. Coloring is a well-known problem in graph theory. However, the assignment problem in CoTDMA does not directly map to the classical coloring problem in graph theory. For CoTDMA, a *modified* construction of the conflict graph, as well as a *modified coloring problem*, is needed to reflect the specifics of 802.11 CSMA scheme, as detailed below.

Instead of pre-assigning the three frequency channels as in Fig. 8, let us first set up a general framework which integrates the frequency channel assignment and time-slot assignment. In the CoTDMA conflict-graph model, we use two layers of coloring. The first-layer colors represent frequency channels and the second-layer colors represent time slots. The first-layer coloring is applied at the cell level (assuming all nodes within the cell use the same frequency channel in a static manner) while the second layer coloring is applied at the station (vertex) level. In the following, we first state the constraints of our coloring problem under the context of 802.11, and then describe how to capture the constraints in the conflict graph. Vertices are associated with the client stations in the following:

*Constraint 1*: The number of available first-layer colors is $m$, and the number of available second-layer colors is $n$ (see Definition 1)

*Constraint 2*: All vertices associated with the same AP (within the same cell) must have the same first-layer color.

*Constraint 3*: Within a cell, there can be at most $k$ vertices assigned with the same second-layer color. Furthermore, the vertices assigned the same second-layer color in a cell must be within the *CSRange* of each other.

For vertices within the same cell, it is obvious that the associated stations cannot transmit together since one end of the links is always the AP. The issue for vertices within the same cell is that whether the *CSRange* of vertices (clients) can cover each other, hence Constraint 3 – i.e., if two vertices are not within each other's *CSRange*, then carrier sensing between them does not work, and therefore they should be assigned different time slots to avoid the hidden-node phenomenon.

*Constraint 4*: Consider two vertices of different cells, $v_i$ and $v_j$. They conflict with each other and must be as-signed different (first-layer color, second-layer color) combinations if one or more of the following inequalities below are satisfied:

$$CSRange \geq \min(d_{v_i,v_j}, d_{v_i,v_j'}, d_{v_i',v_j}, d_{v_i',v_j'}) \quad (5)$$

$$\begin{aligned} IR_{v_i} &> \min(d_{v_i,v_j}, d_{v_i,v_j'}) \\ IR_{v_i'} &> \min(d_{v_i',v_j}, d_{v_i',v_j'}) \\ IR_{v_j} &> \min(d_{v_i,v_j}, d_{v_i',v_j}) \\ IR_{v_j'} &> \min(d_{v_i,v_j'}, d_{v_i',v_j'}) \end{aligned} \quad (6)$$

where $v_i'$ and $v_j'$ are the corresponding APs that $v_i$ and $v_j$ associated respectively. Note that both (5) and (6) describe the conditions under which simultaneous transmissions are not possible. However, there is a subtlety. Inequalities (6) capture the conditions that will lead to collisions. Inequality (5), on the other hand, only says that the CSMA mechanism will prevent the stations from transmitting together – that is, strictly speaking, if (5) is satisfied, the stations could still be assigned the same color combination, and the CSMA mechanism simply prevents simultaneous transmissions (if we did that, Constraint 3 would need to be modified to encompass the overall network). Constraint 4, however, disallows that as a design choice to simplify things. The reasons are as follows: (i) If different color combinations are used whenever (5) is satisfied for two vertices in different cells, then CSMA in different cells will then be decoupled in each of the time slots under CoTDMA, obviating the need for inter-cell CSMA. (ii) When inequalities (5) and (6) are imposed on CoTDMA coloring, we may decrease *CSRange* to only meet the need of intra-cell CSMA, and there is no need for *CSRange* to be large enough to ensure proper CSMA operation across cells. This has the advantage of reducing exposed nodes (EN) across cells [12], hence increasing spatial re-use. Later in this section, we will explore the "optimal" value for *CSRange* through simulations.

*Capturing Constraints 2 to 4 in Conflict Graph*

To capture Constraint 2, we could assign an *AP_ID* to the vertices in accordance with the APs to which they associate. Vertices with the same *AP_ID* must be given the same first-layer color.

For Constraints 3 and 4, an edge between two vertices means they must be assigned different (first-layer color, second-layer color) combination.

To capture constraint 3, there is an edge between two vertices $v_i$ and $v_j$ of the same cell if

$$d_{v_i,v_j} > CSRange \quad (7)$$

Two vertices that are within the *CSRange* of each other could be assigned the same or different second-layer color. However, there can be at most $k$ vertices with the same second-layer color within a cell.

Constraint 4 can be captured by drawing an edge between two vertices of different *AP_ID* if there is a conflict relationship under inequalities (5) and (6). To avoid confusion, it is worth emphasizing again that between vertices of different *AP_ID* (from different cells), we use (5) and (6) for the edge formation criteria. For vertices of the same cell, we use (7) for the edge formation criteria.

A formulation of the CoTDMA problem is as follows:

*2-Color Assignment Problem:* Assign (first-layer color, second-layer color) to the vertices subject to constraints 1 to 4 such that the total number of vertices successfully colored is maximized.

Fig.9 illustrates the idea of CoTDMA. APs (triangles) are at the center of the cells, client stations (circles) in the same cell are associated with the same AP. The solid-line cells use the same frequency channel while the dotted-line cells use the other two frequency channels. For simplicity, we assume the standard 3-frequency-channel assignment here for the frequency channel assignment in CoTDMA. It is worth noting that this standard 3-frequency-channel assignment is not a must for CoTDMA. In this case, we only draw a conflict graph for second-layer coloring. Due to the standard 3-frequency-channel assignment, the vertices represent clients N1 to N6 all have the same first-layer color. Here we assume 802.11b and $n = 6$ (which implies $k = 2$ given a capacity of 12 VoIP sessions per cell), and $CSRange = d_{max}$.

Fig. 10 shows the corresponding conflict graph together with the coloring. The bold numbers in parentheses beside the vertices are the first-layer colors while the other numbers are the second-layer colors. According to Constraint 3, for second-layer coloring, N1 and N4 are assigned COLOR1, N2 and N5 are assigned COLOR2, and N6 are assigned COLOR3. N3 and N5 may interfere with each other under (5) and (6), so an edge is drawn between them. According to Constraint 4, the two vertices must be assigned with different color (first-layer color, second-layer color): in Fig. 10, COLOR1 is assigned to N3 and COLOR2 is assigned to N5.

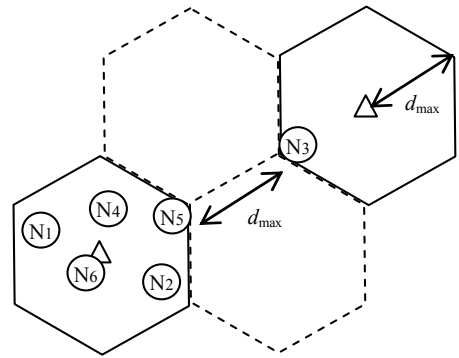

Fig. 9. A topology of VoIP over multiple WLANs with 3-channel assignment.

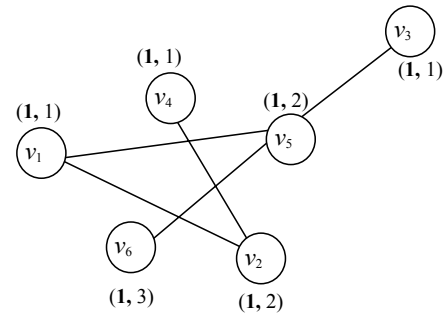

Fig. 10. A conflict graph with first-layer and second-layer colors for the network in Fig. 9.

### 4.1.3 Parameter Values in CoTDMA

An important parameter in CoTDMA is *n*, the number of the time slots (second-layer colors) available for the system. From Definition 1, *k*, the number of VoIP sessions that can be active in the same time slot in a cell is set accordingly to *n*. In the above example, we set $n = 6$ and $k = 2$. The value of *n* directly affects the number of vertices that can be successfully colored. A larger *n* (i.e. a smaller *k*) means more finely-divided time slots. In the extreme case, $k = 1$ ($n = C_{AP\_1}$, i.e. $n = 12$ for 802.11b and $n = 60$ for 11g), which is a pure TDMA scheme. In this case, every VoIP session in a cell is assigned a distinct time slot. Hence, no carrier sensing is required for accessing the medium. From the graph-theoretic coloring viewpoint, fine TDMA as such would allow us to increase the number of vertices successfully colored in the two-layer-color assignment problem defined above. However, fine TDMA has an implementation cost not captured in the coloring problem – namely, there is the need for a "guard time" as we switch from slot to slot. This implementation issue will be further discussed in the next subsection. For the time being, suffice it to say that we are interested in making *k* as large as possible (i.e. making *n* as small as possible) while retaining the performance results of the case where *k* is set to 1.

Another important parameter in CoTDMA is *CSRange*. As mentioned in the explanation of Constraint 4 in Subsection 4.1.2, CoTDMA allows us to decrease *CSRange* and the carrier sensing mechanism needs only to work properly within a cell. However, if the *CSRange* is too small, a client may only carrier-sense few other client stations within the cell. According to constraint 3 (and inequality (7)), small *CSRange* may cause more intra-cell edges in the conflict graph that restrict coloring freedom.

For illustration, let us first consider setting *CSRange* to be the sector diameter, as described below. Each cell is divided into *n* sectors in a fashion that generalizes Fig. 8. *CSRange* could be set to cover the maximum distance between any two points within a sector. Let us refer to the maximum distance as the "diameter" of the sector. For example, if 802.11b is assumed, $C_{AP\_1} = 12$. For $n = 12, 6, 4,$ 3, 2, and 1, the corresponding *CSRange* could be set as shown in Table 3. In Table 3, we also show the corresponding values of *k* according to Definition 1. If 802.11g is assumed, the same set of *CSRange* values should be used if $n = 60, 6, 4, 3, 2$ and 1 (as in 802.11g, $C_{AP\_1} = 60$).

The sector-diameter is a "worst-case" setting in the sense that it assumes that there exist actually two stations at the extreme corners of a sector that define the diameter, which is rarely the case. It would therefore be of interest to explore tuning the *CSRange* to maximize the number of vertices can be successfully colored in the conflict graph. We present the simulation results based on the sector-diameter settings as well as "smaller-than-sector-diameter" settings below.

We first show the results of sector-diameter settings. Based on the parameters in Table 3, we have performed MATLAB experiments assuming 802.11b ($C_{AP\_1} = 12$) and 802.11g ($C_{AP\_1} = 60$). For simplicity, the standard 3-frequency-channel assignment is assumed. Since the first-layer colors are pre-fixed in this experiment, only second-layer coloring (time-slot assignment) is considered. We use hexagonal cells to model WLAN, and 12 (for 802.11b) or 60 (for 802.11g) wireless client stations are randomly placed inside each cell with uniform distribution.

We use a heuristic algorithm of Welsh and Powell [24] to color the conflict graph. We add our coloring constraints 1 to 4 to tailor the algorithm to CoTMDA. The algorithm of Welsh and Powell does not give optimal graph coloring in general, but the effect of *n* is already quite pronounced even with the simple heuristic.

Fig. 11 and 12 respectively show the percentages of vertices that can be colored when $C_{AP\_1} = 12$ and $C_{AP\_1} = 60$. In the experiment, *CSRange* changes accordingly to *n* as dictated by Table 3. For each *n*, we ran five experiments with different node distributions. In the figures, circles are the average percentage of successfully colored vertices. In general, as *n* increases, more vertices can be successfully colored. Although not shown in the figures, the two cases for $n = 12$ in 802.11b and $n = 60$ in 803.11g have 100% of their vertices successfully colored in all runs of our experiments. But even for *n* as small as 3 or 4, the performance of CoTDMA is already very good (100% of vertices are colored in 802.11b, while more than 95% of vertices are colored in 802.11g). With smaller *n*, the overhead of guard-time for switching between time slots can be reduced (to be elaborated shortly) and $n = 3$ or 4 may offer the best design tradeoff.

TABLE 3
VALUE OF *n*, *k* AND CORRESPONDING *CSRange* IN 802.11b

| *n* | 1 | 2 | 3 | 4 | 6 | 12 |
|---|---|---|---|---|---|---|
| *k* | 12 | 6 | 4 | 3 | 2 | 1 |
| *CSRange* | $2d_{max}$ | $\frac{\sqrt{13}}{2}d_{max}$ | $\sqrt{3}d_{max}$ | $\frac{\sqrt{7}}{2}d_{max}$ | $d_{max}$ | 0 |

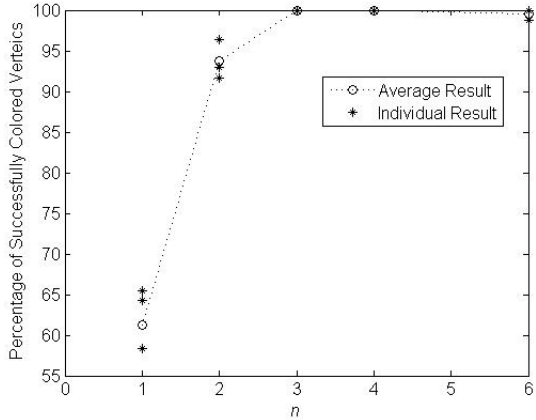

Fig. 11. Percentage of colored vertices when *CSRange* is set as "sector-diameter" and $C_{AP\_1}$=12 (802.11b).

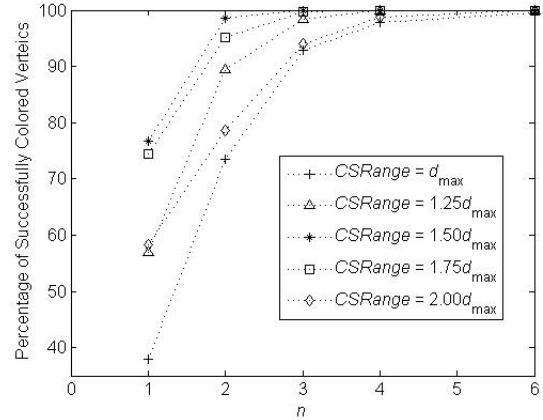

Fig. 13. Average percentage of colored vertices as *CSRange* changes when $C_{AP\_1}$=12 (802.11b).

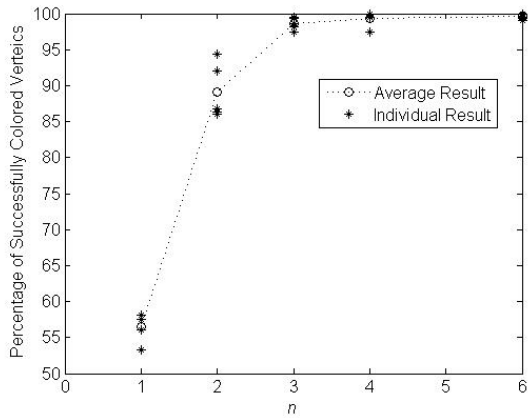

Fig. 12. Percentage of colored vertices when *CSRange* is set as "sector-diameter" and $C_{AP\_1}$=60 (802.11g).

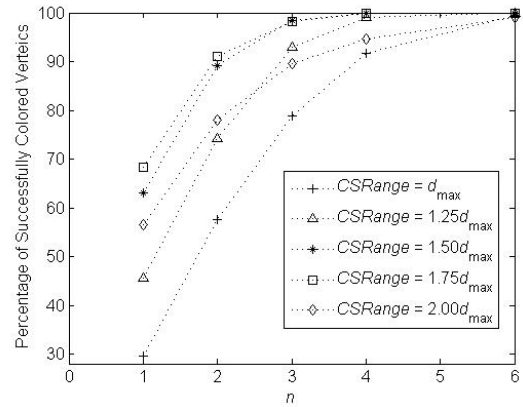

Fig. 14. Average percentage of colored vertices as *CSRange* changes when $C_{AP\_1}$=60 (802.11g).

In the next set of experiments, for each run, we set a fixed *CSRange* for all values of *n*. Across different runs, we vary *CSRange*. Fig. 13 and 14 show the average percentage of successfully colored vertices as a function of *n* under different fixed *CSRange*. We find that when *CSRange* is $d_{max}$, the overall system performance is poor because many edges are formed within a cell (according to (7)).

As *CSRange* increases, the overall system performance improves. But beyond certain point (e.g., $1.5d_{max}$ in the figures), the overall system performance drops again. This phenomenon reveals the tradeoff between intra-cell optimality and inter-cell optimality. When *CSRange* is too large, say $2d_{max}$, many edges are formed between vertices of different cells (according to (5)), leading to an increase of exposed nodes. Through experimentation, we find the "optimal" *CSRange* which yields the "best" system performance is around $1.637d_{max}$. Indeed, we could use this setting for different *n* values with reasonably good results (see Table 4).

Our experiment results show that even though the freedom of the two dimensions of frequency and time in CoTDMA has not been fully utilized (due to our assuming the fixed 3-frequency-channel assignment), CoTDMA can generally color large portions (over 90%) of the vertices of the conflict graph when $n \geq 3$ for both 802.11b and 802.11g. The performance can be even better when appropriate ("optimal") *CSRange* is set. By contrast, without CoTDMA, and with 3-frequency-channel assignment alone (*n* = 1), we find that only around 60% of the total capacity can be utilized.

TABLE 4
SYSTEM PERFORMANCE FOR $CSRange$ = $1.637d_{max}$

| n | 1 | 2 | 3 | 4 | 6 | $C_{AP\_1}$ |
|---|---|---|---|---|---|---|
| % vertices successfully colored in 802.11b | 80.0 | 98.6 | 100 | 100 | 100 | 100 |
| % vertices successfully colored in 802.11g | 69.8 | 93.0 | 99.6 | 100 | 100 | 100 |

## 4.2 Possible Realizations of Time Division Multiple Access within Existing IEEE 802.11 Standards

Most previous work [18-23] considered proprietary protocols for implementing TDMA on wireless networks. Our focus here is to implement time-slot assignment within the framework of the IEEE 802.11 standard. A most critical issue is how to realize the concept of "time slot" within the 802.11 CSMA structure. A possibility is to make use of the "sleeping mode" in 802.11, which was originally designed for power conservation purposes. In the sleeping mode, beacon frames are used for synchronization. Accordingly, in CoTDMA, all stations could wake up around the beacon time for synchronization. As illustrated in Fig. 15, in CoTDMA, within each beacon interval ($BI$) between the *end* of a beacon and the *beginning* of the next beacon, the time is divided to $C$ *frames* (cycles), each of duration $\Delta T$. Within each frame, there are $n$ *time slots*, each of duration $\Delta t$. A VoIP stream transmits *one* VoIP packet in each frame. The offset from the end of the beacon to the beginning of the $i^{th}$ frame is $(i-1)\Delta T$. A station that has been assigned time slot $j$ is to be awake within a $BI$ only during the time intervals specified by [ $(i-1)\Delta T + (j-1)\Delta t$, $(i-1)\Delta T + j\Delta t$ ), $i$ = 1, 2, ..., $C$. Other than these time intervals and the beacon time, the station sleeps.

**Guard-Time Overhead in CoTDMA**

The guard time should be set to the duration of one VoIP packet. That is, no packet transmission should be initiated within the current time slot when the beginning of the next time slot is only a guard time away. This is to ensure packet transmission will not straddle across two time slots. Let $r_{\Delta t}$ be the *maximum* number of VoIP packets (including the CSMA overhead) that could be transmitted and received within each time slot by all VoIP sessions which are active in that particular time slot. Then,

$$r_{\Delta t} = 2 \times \frac{R_{VoIP}}{10nC} \times C_{AP\_1} \qquad (8)$$

where $R_{VoIP}$ is the number of VoIP packets generated per second in a particular VoIP codec. The factor of 2 is due to each VoIP session having a downstream and an upstream flow. The factor of $1/10$ is due to the default 0.1s separation time between two beacons. The guard time overhead is a constant of one VoIP-packet duration, so that the time-slot efficiency is $(r_{\Delta t} - 1)/r_{\Delta t}$. Thus, smaller $r_{\Delta t}$ gives rise to lower efficiency, which in turn results in lower capacity. From (8), we can see that it is desirable to set $n$ and $C$ as small as possible. From the simulation experiments in the previous subsection, however, we need to make sure that $n \geq 3$ (see Subsection 4.1) so as to make sure most sessions within the system capacity limit can be admitted. While call admission consideration imposes a limit on how small $n$ can be, the delay budget consideration imposes a limit on how small $C$ can be, as explained in the next few paragraphs. In other words, the factors that bound on the size of $n$ and $C$ are different.

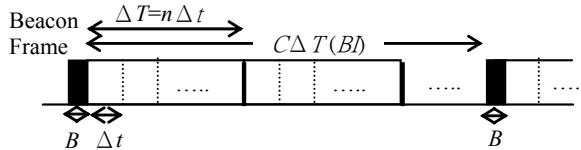

Fig. 15. Frame structure of CoTDMA when it is implemented using 802.11 sleeping mode.

In CoTDMA, the maximum delay for a VoIP packet is $\Delta T + B$, where $B$ is the duration of the beacon (see Fig. 15). To see this maximum delay, let us consider the station being assigned time slot $j$. In the worst case, it could generate a packet in the last frame within a $BI$ just slightly after time slot $j$ ends, thus missing it. The earliest time for the next time slot $j$ (in the next $BI$) is $\Delta T + B - \Delta t$ after that. Within this next time slot $j$, in the worst case, the packet is sent just before the end of the timeslot (due to the CSMA contention with other stations assigned the same time slot). If we assume the packet will be discarded if it fails to be sent out by this time - so as to make way for a newly generated packet from the same VoIP session - the maximum delay is then $\Delta T + B$. Thus,

$$\Delta T + B \leq DB \qquad (9)$$

where $DB$ is the delay budget. Since $C\Delta T = BI$ (see Fig. 15), we have

$$BI/C + B \leq DB \Rightarrow C \geq \frac{BI}{DB - B} \qquad (10)$$

Suppose we set a local delay budget of 30ms for VoIP applications [1]. A typical value of $B$ is 0.5ms. With the default separation time between two beacons of 100ms, a beacon interval, $BI$, is 99.5ms. Hence, the number of frames, $C$, in a beacon interval, is at least 99.5/29.5 = 3.37. $C$ should be a positive integer as well, so the smallest $C$ is 4.

Assuming $n$ = 3, $C$ = 4, and GSM 6.10 codec and 802.11b ($R_{VoIP}$ = 50, $C_{AP\_1}$ = 12), from (8) we find that 90% of capacity is utilized. That is at most $\lfloor 12 \times 90\% \rfloor = 10$ VoIP sessions can be admitted per AP in 802.11b networks. If 802.11g is assumed ($C_{AP\_1}$ = 60), 98% of capacity is utilized. That is 58 VoIP sessions can be admitted per AP in 802.11g networks.

Take 802.11b WLAN. With 3-frequency-channel CoTDMA, the per-AP capacity over multiple WLANs is 10 VoIP sessions. This is a 35.3% improvement over the per-AP capacity of 7.39 sessions for the three-frequency-channel clique-analytical call admission control strategy in Section 3.3.

Another possibility for implementing the concept of time-division is to use the "polling mechanism" of Point Coordination Function (PCF) [25] to imitate the time slot assignment in CoTDMA. In PCF, traffic is scheduled by the AP, so that no extra guard-time overhead is needed for time slot switching. If we assume PCF, we could set $n$ = $C_{AP\_1}$ (i.e., $k$ = 1) in our CoTDMA call admission scheme, so that we essentially have the extreme case of Fine-grained Time Division Multiple Access (FiTDMA). In PCF, an AP maintains a polling list containing all the wireless stations in its WLAN. In the contention-free period, the AP polls the stations on the polling list one by one, since only one VoIP session is active in each timeslot in a cell in FiTDMA. Only the polled station can transmit packets to AP (the downstream packet is attached in the polling packet). The position of a station in the polling list corresponds to the time slot assigned to the station. In this way, stations which may interfere with each other in adjacent cells will not be polled at the same time.

With FiTDMA, the number of vertices successfully colored can be increased. Furthermore, since only one VoIP session is active in each timeslot in a cell, carrier sensing mechanism can be deactivated. Without backoff countdown in contention period, $C_{AP\_1}$, the average per-AP capacity in single isolated WLAN, can be boosted in PCF (in 802.11b, $C_{AP\_1}$ increases from 12 to 17, while in 802.11g, $C_{AP\_1}$ increases from 60 to over 90). A major concern, however, is that PCF is seldom used in practice and many 802.11 devices do not support it – unlike DCF, the robustness of PFC in field deployment has not been well tested.

## 5 CONCLUSION

In this paper, we have shown that when there are multiple 802.11 WLANs within the vicinity of each other, the already low VoIP capacity in the single isolated WLAN case (around 12 and 60 VoIP sessions per AP in 802.11b and 802.11g, respectively) is further eroded in a very significant manner, so much so that less than 1% goodput can be supported. For example, in 802.11b, less than 2 VoIP sessions per AP can be supported while throughput computation based on raw bandwidth of the WLAN yields 200 sessions per AP. In 802.11g, around 10 VoIP sessions per AP can be supported while 900 sessions can be yielded from throughput computation. The dismal performance and inefficiency imply that there is much room for improvement within the 802.11 standard as far as the support for VoIP is concerned.

The low VoIP capacity in the single WLAN [1, 2] is due largely to header overheads, and packet aggregation [1, 2] is an effective solution to reduce the header penalty. The further degradation of VoIP capacity in the multiple-WLAN case, however, is due to interferences among the WLANs, and requires additional solutions. This paper suggests a two-pronged approach **1) call admission control**; and **2) virtual channelization**.

Regarding 1, we have formulated a clique-analytical call admission control algorithm, and shown that (compared with a simplistic call admission scheme) it can improve the VoIP capacity in a 5-by-5, 25-cell 802.11b WLAN by 52.1% from 1.63 sessions to 2.48 sessions per AP. The improvement is 36.8% in 11g. If three orthogonal frequency channels are used, such as are available in 802.11b/g, the capacity can be increased to 7.39 (11b) and 44.91 (11g) VoIP sessions per AP by careful frequency-channel assignment to the cells.

Regarding 2, the three orthogonal frequency channels in 802.11b/g are not enough to completely isolate interferences among WLANs. This in turn requires the CSRange of 802.11 to be set to be rather large to prevent packet collisions; but doing so also increase the exposed-node problem that degrades the VoIP capacity. In this paper, we have shown that "virtual channels" (or timeslot channels) could be created to combat this problem. By assigning the virtual channels judiciously to the VoIP stations, we could effectively isolate the interferences between cells. Specifically, we have proposed coarse-grained time-division multiple access (CoTDMA) as a means for virtual channelization within the basic 802.11 CSMA protocol, so as to maintain compatibility with the widely-deployed 802.11 technology to a large extent. In CoTDMA, the time dimension is divided into multiple coarse time slots; multiple VoIP sessions are assigned to each time slot, and the VoIP sessions assigned to the same time slot in a WLAN make use of the basic 802.11 CSMA protocol to coordinate channel access. The basic idea is that VoIP sessions of adjacent WLANs that may interfere with each other in a detrimental way should be assigned different time slots or frequency channels, so that VoIP sessions of different cells do not need to use CSMA to coordinate transmissions among them, and that CSMA is in effective use only within a cell.

We note that CoTDMA is fundamentally a technique in which stations contending for a common resource (i.e., airtime) are compartmentalized into subsets so that only stations within each subset contend with each other. The partitioning is done in such a way that the subnetwork consisting of stations within each subset is less susceptible to detrimental interference/carrier-sensing pattern so that the resource could be used more efficiently. This principle is applicable not just to voice traffic, but to wireless networking in general with or without voice traffic. As further work, we note that the graph-theoretic formulation of the two-layer coloring problem in Section 4 is rather general, and could be explored for the general case.

In the simulation experiments of this paper, even though the freedom of the two dimensions of frequency and time in CoTDMA have not been fully exploited (due to the assumption of a standard 3-frequency channel assignment to the WLANs), we find that CoTDMA could already improve the VoIP capacity meaningfully. Our results indicate that with a small number (3 to 4) of coarse time slots in CoTDMA, the per-AP VoIP capacity can be increased to 10 sessions in 802.11b and 58 sessions in 802.11g (another 35.3% and 29.15% improvement over clique-analytical admission control with three orthogonal frequency channels, respectively).

Last but not least, we note that although the effect of mutual interference on VoIP over *multi-hop mesh networks* has also been considered [7-9], our proposed CoTDMA scheme is targeted at solving mutual interference in *infrastructure networks*, the predominant architecture deployed in the field today. Having said that, it is still interesting to explore whether and how similar approaches could be applied to multi-hop networks.

**An Chan** received the B.Eng and M.Phil degrees in Information Engineering from The Chinese University of Hong Kong, Hong Kong in 2005 and 2007 respectively. He is currently working toward a Ph.D degree in the Department of Computer Science at the University of California, Davis. His research interests are in QoS over wireless network and advanced IEEE 802.11-like multi-access protocols. He is a graduate student member of IEEE.

**Soung Chang Liew** received his S.B., S.M., E.E., and Ph.D. degrees from the Massachusetts Institute of Technology. From March 1988 to July 1993, Soung was at Bellcore (now Telcordia), New Jersey, where he engaged in Broadband Network Research. Soung is currently Professor and Chairman of the Department of Information Engineering, the Chinese University of Hong Kong. Soung's current research interests focus on wireless networking. Recently, Soung and his student won the best paper awards in the *1st IEEE International Conference on Mobile Ad-hoc and Sensor Systems (IEEE MASS 2004)* the *4th IEEE International Workshop on Wireless Local Network (IEEE WLN 2004)*. Separately, TCP Veno, a version of TCP to improve its performance over wireless networks proposed by Soung and his student, has been incorporated into a recent release of Linux OS. Publications of Soung can be found in www.ie.cuhk.edu.hk/soung.